# COUNTERFACTUAL ANALYSIS BY ALGORITHMIC COMPLEXITY: A METRIC BETWEEN POSSIBLE WORLDS

**NICHOLAS CORRÊA**
https://orcid.org/0000-0002-5633-6094
*PUCRS*
*Graduate Program in Philosophy*
*Department of Philosophy*
*Porto Alegre, R.S.*
*Brazil*
nicholas.correa@acad.pucrs.br

**NYTHAMAR FERNANDES DE OLIVEIRA**
https://orcid.org/0000-0001-9241-1031
*PUCRS*
*Graduate Program in Philosophy*
*Department of Philosophy*
*Porto Alegre, R.S.*
*Brazil*
nythamar.oliveira@pucrs.br

**Abstract:** Counterfactuals have become an important area of interdisciplinary interest, especially in logic, philosophy of language, epistemology, metaphysics, psychology, decision theory, and even artificial intelligence. In this study, we propose a new form of analysis for counterfactuals: analysis by algorithmic complexity. Inspired by Lewis-Stalnaker's

Possible Worlds Semantics, the proposed method allows for a new interpretation of the debate between David Lewis and Robert Stalnaker regarding the Limit and Singularity assumptions. Besides other results, we offer a new way to answer the problems raised by Goodman and Quine regarding vagueness, context-dependence, and the non-monotonicity of counterfactuals. Engaging in a dialogue with literature, this study will seek to bring new insights and tools to this debate. We hope our method of analysis can make counterfactuals more understandable in an intuitively plausible way, and a philosophically justifiable manner, aligned with the way we usually think about counterfactual propositions and our imaginative reasoning.

## 1. Introduction

A counterfactual expresses something contrary to the fact, that is, something that is not the case. Something that didn't happen. This linguistic artifact involves our intuitive notion that the human mind is capable of conjecturing alternatives contrary to what happened. Our imaginative reasoning is constantly employed to evaluate what could have happened differently ('If it were this A, then it would be this B'). For example, "if I had come home that way, it would have taken me less time." Counterfactuals have been of interest to philosophers since the time of Leibniz in the 17th century, where the German philosopher proposed the possibility of an infinite number of alternative realities [Griffin 1999], that is, infinite possible ways in which real-world events could have evolved differently.

In modern literature, counterfactuality is an object of interest in several areas:

- in social psychology, counterfactual thinking is theorized to serve as a behavioral regulator [Epstude and Roese 2008];
- in the cognitive sciences, the imagination of alternative realities (rational imagination) is believed to be intrinsically linked to rational thinking, the formation of rational intentions, and human reasoning in general [Byrne 2008; 2016];
- in the study of decision making, an extremely interdisciplinary area of research, various proposals of Decision Theory correspond to different ways of formalizing counterfactual reasoning [Roese 1999];
- research in artificial intelligence has a great interest in these objects, being important for formalizing how autonomous agents can make their decisions optimally [Ginsberg 1985; Pearl 1995; Costello and McCarthy 1999; Bottou et al 2013].

Similarly, the literature involving formal semantics, philosophy of language, logic, epistemology, and metaphysics has various interpretations of how counterfactuals should be understood and analyzed, something that makes these objects a great source of curious questions, with many possible interpretations. After all:

- How can we express, in a way that satisfies our intuitions, the form we reason and conjecture about counterfactual possibilities?

In an attempt to answer this question, we would like to propose a different way of thinking about counterfactuals and introduce new tools to analyze these objects. This study aims at formalizing a similarity function between possible worlds to extend the work of David Lewis [1973a], Robert Stalnaker [1968], and Todd William [1964], and their Possible Worlds Semantics. We hope to bring in a new perspective and thus a new vision for old problems and disputes.

## 2. Goodman's problems

The term counterfactual, coined by Nelson Goodman [1947], is a more succinct way of expressing a *condition contrary to facts*, popularized by Roderick Chisholm. Counterfactuals, within the study of formal language, are a class of conditional sentences that allow their speakers to discuss about possibilities, such as 'If it weren't raining now, I would be on the beach.' In this way, we can define counterfactuals as necessarily having a false antecedent, as opposed to indicative conditionals which may have false or true antecedents [Kaufmann 2005]. From this realization, we arrive at our first problem, i.e., the vagueness that these objects express.

When dealing with conditional natural language[1], statements like 'if $P$ then $Q$' are true whenever the antecedent $P$ is false. Since counterfactual statements have precedents that are by necessity false, this would imply that all counterfactual statements are true, or vaguely true. This impossibility of expressing counterfactuals within classical logic has prompted philosophers such as Willard van Orman Quine [1960; 1982] to claim that counterfactuals are not logical, and therefore make neither false nor true statements about the real world.

Goodman [1947] illustrates this point using two well-known examples, where both, no matter how much they state opposing propositions, would be equally true:

[1] Even though conditional natural language statements do not necessarily imply a material implication.

a) If that piece of butter had been heated to 150 °, it would have melted.
b) If this piece of butter had been heated to 150 °, it would not have melted.

The second problem involves the context-dependence of counterfactuals. Quine [1960; 1982] states that of the following two statements, it is impossible to determine which would be *truer* than the other (reiterating that both could not be true at the same time):

c) If Caesar had been in command of Korea, he would have used the atomic bomb.
d) If Caesar had been in command of Korea, he would have used catapults.

For Quine, the fact that one cannot define which of the propositions would be the true one shows that counterfactuals are not connected to real-world states but rather to the imagination and purpose of the speaker.
Finally, counterfactuals are non-monotonic, insofar as their truth value can be changed as we add extra material to their background. Natural language conditioners cannot possess this property, known as the principle of Antecedent Strengthening. Goodman [1947] was one of the first to report that the non-monotonicity of counterfactuals can turn a true counterfactual proposition into a false one, like in this example:

e) If I had struck this match, it would have lit.
f) If I had struck this match and done so in a room without oxygen, it would have lit.

Examples (e) and (f) show that the truth condition of a counterfactual depends on assumed facts (presence of oxygen), which makes the first example true, and the second false. Goodman argues that it is quite difficult to specify all the details since numerous factors can be added (wind, rain, fake matches, an invisible entity who keeps blowing off the flame produced by the speaker). One could say that '*truer*' counterfactuals are those that respect the laws of physics of our universe (or the universe of the speaker). For Goodman, to adequately specify all the background factors, together with all relevant physical laws, would be quite complex in non-counterfactual semantic terms.

However, instead of thinking of these points as problems, we can define them as the *very* characteristics that define counterfactuals:

- *Vagueness:* counterfactuals are not apt for true statements, being at most vaguely true;
- *Context-Dependence:* Counterfactuals have context-sensitive truth conditions;
- *Non-monotonicity:* Counterfactuals must be interpreted in a non-monotonic way.

Thus, any theory that seeks to analyze and formalize these objects must cope with such demands. A popular theory that seeks such formalization, however, not without its critics, is the Possible world semantics.

## 3. Possible world semantics

The use of modal logic and Kripke's notation allowed counterfactuals to be understood within an appropriate formal language. However, the use of modal logic to express counterfactual conditionals has led to different theories and interpretations within the academic community. Currently, theorists are divided on how best to formalize and understand counterfactuals within the Possible world semantics framework. And as we will see, there are still those who invalidate it as a method of analysis.

The Possible world semantics, also known as similarity analysis, tries to analyze in a logically valid way counterfactual propositions respecting their characteristics, such as vagueness, context-dependence, and non-monotonicity. One of the main concepts of this method is the concept of a possible world. We define as a possible world ($w$), a way in which the real world could have been, being part of a set of possible worlds $w \in W$. It is the proximity relation of a possible world to the real world that attributes the truth condition to a counterfactual statement ($\phi$), where $\phi$ can be considered true in a possible world $w$, provided that certain conditions of similarity between the real/present world and the possible world are satisfied [Kripke, 1963]. Lewis [1973b] proposes that the non-monotonic characteristic of counterfactuals can be formalized within the similarity analysis through a system of possible worlds nested by similarity. Think of a sphere, with several other concentric spheres around it, the central one ($w_o$) being a singleton containing the actual/real world. As possible worlds distance themselves from this singleton, we transit to spheres that may contain many other possible worlds. Worlds where a counterfactual can have a different truth value.

Thus, possible worlds ($w_i$) are ordered by their similarity to the real world, $w_o$, where the most similar to $w_o$, where the antecedent ($\phi$) is true, are possible worlds where the consequent ($\psi$) is also true. We can formalize this method as follows:

- there is a similarity function $f$, which takes as its input a $w$-world, a $\phi$ -world (expressed by a

> proposition $\phi$), and returns the set of $\psi$ -worlds most similar to $w$-worlds. Thus, $\phi > \psi$, where '>' is the logical connective, is then said to be true when the $\phi$-worlds most similar to $w_0$ are $\psi$-worlds.

Within the formalism suggested by the similarity analysis, there are divergent interpretations of what is known as the *Limit Assumption* and the *Singularity Assumption*. For Stalnaker, there is a continuous chain of possible worlds, each one closer to the real world (Limit Assumption). However, there is only one possible world that is closer to the real world (Singularity Assumption). That is to say, $\phi$-worlds don't become indefinitely more similar to $w_0$ [Lewis 1973b: 77 - 83]. The Limit Assumption also states that this chain has minimal elements, '<' being the predicate that defines a chain. A chain where there is at least one $w_i$ such that there is no $w_j$ with $w_j < w_i$. The singularity assumption, then, states that $w_i$ is unique.

Lewis rejects the Limit and Singularity assumptions, arguing in favor of the idea that there may be possible worlds that come closer and closer to the real world, continuously and without limit.

Stalnaker [1980], in defence of the Singularity assumption, proposes the Law of excluded middle, which dictates that all instances of formulas $(\phi > \psi) \vee (\phi > \neg\psi)$ are true. According to the Singularity assumption, for each antecedent $\phi$, there is only one possible world closer to where $\phi$ is true. In turn, the Law of the excluded middle dictates that any consequent $\psi$ is true, or false, in that singular world where $\phi$ is true. For example:

a) If the fair coin had been tossed, it would have resulted in heads.

b) If the fair coin had been tossed, it would have resulted in tails.

For Stalnaker, there is only one possible world closer to $w_0$ where the coin falls either head or tails, making (a) true and (b) false, or (a) false and (b) true. Lewis [1973b], on the other hand, maintains that both (a) and (b) are false, for there are no nearest possible worlds where the coin falls either heads or tails.
Possible worlds semantics, or another form of similarity analysis, depends on the restrictions imposed on the similarity function $f$. As we can see above, the functions of Stalnaker and Lewis differ on points related to the Limit and Singularity assumptions. For Lewis (Lewis 1973b: 91 - 96), $\phi > \psi$ is vaguely true if, and only if:

- there are no worlds where $\phi$ is true ($\phi$ is metaphysically impossible).

And not vaguely true if, and only if:

- between the worlds where $\phi$ is true, and some worlds where $\psi$ is true are closer to the real world

  than any possible world where $\psi$ is false, or $\phi > \psi$ is false otherwise.

However, one of the most frequent criticisms made to both Stalnaker and Lewis' similarity analysis is the vagueness with which the authors define the similarity function. In the words of Lewis [1973: 92], the similarity function $f$ is described as: "[...] *our familiar and intuitive concept of comparative global similarity* [...]". This is not a very strict (or formal) notion for expressing a similarity function. But Lewis is clear in his work, stating that his notion of proximity is only an intuitive notion and not a metric of proximity. Lewis [1979] sought to further formalize his similarity function by establishing the following weight system, which would define rules for establishing the similarity between possible worlds:

- Avoid great miracles, that is, violations of physical laws that characterize the real world;
- Maximize the entire space-time region in which the perfect combination of particular facts prevails;
- Maximize the period during which similar worlds coincide in matters of facts;
- Avoid even small miracles;
- Facts that occur after the private facts involved need not be kept fixed.

Nevertheless, the informality as such a weight system presented makes the similarity analysis, at best, an incomplete analysis. Probably there are a large number of possible similarity functions. If the similarity analysis is to be expanded to a more complete theory we need to answer more rigorously the following question:

- On what basis do speakers determine that some possible worlds are closer than others?
- Can we express a similarity function more formally?

## 4. Critiques to Possible Worlds Semantics

First of all, we would like to point out that the literature on counterfactuals is *extremely extensive*. It is not possible to make a fair review of all the existing theories and proposals in a single article. However, other methods of analysis do exist, with their particular tools. For example, Strict Analysis [Warmbrōd 1981a; 1981b; Gillies 2007], Conditional Probability Analysis [Adams 1976; Edginton 2003; 2014], and Structural Equations/Causal Models [Pearl 2013]. Several authors, such as Schulz [2007], Kvart [1986; 1992], McGee [1989], Bennett [2003], Bradley [2002], come to prefer probabilistic analysis, inspired by Adams' pioneering work [1976], while some are openly critical of the possible worlds semantics.

Hannes Leitgeb [2012a, 2012b], another theoretician in favor of probabilistic analysis, states that possible worlds semantics, unlike probabilistic analysis, is not capable of corresponding to any form of magnitude of probability that represents a consistent order of similarity. This is an opinion shared by other critics of the similarity analysis model, such as Hájek and Edgington. In the words of Hájek:

> I have long argued against such similarity accounts. Worlds in which the plate is dropped and falls to the floor may well be more similar to ours than worlds in which it is dropped and does something else. But that doesn't make it true that if the plate were dropped, it would fall to the floor. That counterfactual is undermined by the fact that if the plate were dropped, it would have a positive chance of not falling to the floor [...] This chance is indifferent to how similar is a world where this happens [2014: 250].

Hájek even goes so far as to criticize Lewis' weight system as non-scientific: '*Science has no truck with a notion of similarity; nor does Lewis's [1979] ordering of what matters to similarity have a basis in science* [2014: 250]'. Morreau [2010] also argues that the similarity analysis proposed by Lewis [1979] would not be enough to assess all the differences between possible worlds:

> The trouble comes to light when we ask just how to combine similarities and differences in various respects. In fact, no one has had any real idea! There are only metaphors, however promising these might seem. [...] We cannot add up similarities or weigh them against differences. Nor can we combine them in any other way. Goodman was right to be skeptical. No useful comparisons of overall similarity will result [Morreau 2010: 471].

We argue that there is a way to strengthen the possible worlds semantics, and similarity analysis in general, providing a similarity function that can (2) formalize a distance between possible worlds, (2) proposes a solution to the divergences between Lewis and Stalnaker's interpretations, and (3) that can meet the demands and problems raised by Goodman.

## 5. Defining a Similarity Function: Algorithmic Information and Complexity

In his article *'Why Philosophers Should Care About Computational Complexity',* Aaronson [2013] brings up arguments on how Complexity Theory can aid philosophical investigations involving the nature of knowledge, the problem of logical omniscience, Hume's induction problem, issues involving rationality, among several others. In other areas, complexity theory has already been used in problems involving sequences of random numbers [Kolmogorov 1998], in the definition of methods for inductive inference [Solomonoff 1964], in general, artificial intelligence models [Shane and Hutter 2007], and even to model biological evolution [Chaitin 1991; 2006]. Inspired by this type of interdisciplinary research, which invites the philosopher to study other areas, in the same form that we invite other fields to study philosophy, we propose a new method of similarity analysis using algorithmic complexity as a tool.
Two of the key concepts that will be fundamental to formalize this method are information and complexity. Both concepts are the focus of Algorithmic Information Theory, and it will be through them that we intend to propose a metric between possible worlds. The main insight of algorithmic complexity is that information and complexity are two related concepts. This idea can be understood in the following way:

> A gas takes a large program to say where all its atoms are, but a crystal doesn't take as big a program, because of its regular structure. Entropy and program-size complexity are closely related [...] [Chaitin 2007: 119].

In other words, the information contained within a system, and the complexity of this system, are closely related. The more ordered a system is, the less complex the algorithm needed to produce it will be. To better formalize the idea of Lewis and Stalnaker, more specifically the similarity function $f$, we first need to define two other concepts, the first being *digitalization*.
Digitalization is the process of converting analog information into a digital format, converting analog source information into a sequence (string) of numbers. Digital representations have useful properties, allowing information of all types and in all formats to be transported and processed in a single language, such as the binary alphabet [McQuail 2000: 16 - 34].

In principle, all information can be represented in binary sequences. In our case of interest, we need the environment ($w_0$) of the speaker (agent) to be represented in a digital format. Any amount of information can be represented by a sequence of 0's and 1's, as long as this information is finite. The size of the sequence or the multiplicity of information contained by the environment is irrelevant from a theoretical point of view. Multiple sequences can be concatenated one after the other so that given the correct adjustment in the processing of the sequence, the results will be the same.

A digital image, your voice recorded in a microphone, a video, weather conditions, all can be converted into binary representations of reality. This idea is not new. Leibniz was one of the precursors of binary notation, emphasizing the inexhaustible combinatory potential of 0 and 1 [Bell 2000: 517]. Given enough information, such representations may be enough to capture the general concept of a specific world-state. With world-state, we mean a possible world, limited by the causal relationships affecting the agent. For example: when modeling the counterfactual possibility of a local event, we do not need to digitalize Alpha Centauri, Mars, or the other side of the city.

The second necessary concept is that of a *Turing Machine*. A Turing Machine is an abstract mathematical object, and through its mechanism, any algorithm can be computed [Turing, 1936]. Turing machines can perform any computable process, and a Universal Turing machine can perform any process that any Turing machine can. We start from the physical and metaphysical assumption that the environment can be represented by a Turing machine, known as digital physics/metaphysics [Fredkin 2003; Steinhart, 1998]. Thus, the environment (world-state) can be considered as a Turing machine, where the environment has its internal dynamics (program = the laws of physics), which reads the inputs made available by the agent (actions), and according to the input and its internal state, produces the next world-state. Thus, a world-state is a binary sequence $x$, produced by an environment $w \in W$, where $W$ is the set of all computable environments.

We now have a formal way of talking about the environment and world-states. To define a similarity function between world-states, as proposed by Lewis and Stalnaker, we need to compare the current world-state $x$ and a possible counterfactual world-state $x'$. So, in terms of counterfactual analysis from an agent's point of view, we want to know:

- if the agent *had executed* the action $a$, which would generate the counterfactual the world-state $x'$, what is the difference between $x$ and $x'$? How similar are both states? And how likely are they to be generated by the same environment $w \in W$?

To answer this question, we used a tool of Algorithmic Information Theory, algorithmic complexity [Solomonof 1964; Kolgomorov 1998; Chaitin 2007]. The algorithmic complexity, also known as Solomonoff-Kolmogorov-Chaitin complexity, of a world-state, or any finite binary sequence $x$, is the length of the shorter Turing machine (a program), in this case representing the environment, than it would produce $x$. The insight of this method is that the simplest Turing machine which would produce $x'$, is the environment that produces the closest counterfactual world-state to the real world-state. Informally, we can explain the concept of algorithmic complexity as follows:

Given the two sequences of 60 symbols:

1) 101010101010101010101010101010101010101010101010101010101010
2) cMxG9oNO64ceMaVcRnVSB6u5Au86MffRlNrM4DoB0GsTz9JUniCemcjvfb

Which has the greatest complexity?

Sequence 1 has a predictable structure (1 and 0 repeatedly). However, sequence 2 is random, and it is its own minor description. Therefore, the complexity of 1 < 2. Thus, to evaluate the complexity of world-states we use the function $K(.)$, which takes as input a binary sequence $x$, $K(x)$, and results in the shortest program that would produce $x$:

$$K(x) \coloneqq min_p\{l(p): U(p) = x\}$$

Where $p$ is a binary sequence that we call a program, $l(p)$ is the length of this sequence in bits, and $U$ is a prefix of the universal Turing machine called the reference machine. The algorithmic complexity, in this case, is used as a metric to quantify the similarity between finite sequences of information, in the proposed context, the similarity between world-states $x$ and $x'$. Solomonoff [1964] showed that there is always a machine capable of computing $x$ with the following property:

$$K(x) \leq k + 1 \text{ for all binary sequences } x \text{ of length } k.$$

Since if there is no efficient way to calculate a random binary sequence $x$, we can always include the binary sequence as a table in the program, so we only need to add one bit to the sequence to get a program to perform its calculation, where, for example, $K$ produces the sequence $x$ when given the entry ‘0’. This definition of complexity also allows us to formalize the concept of randomness. A numerical sequence is random if there is no way to compress it to an algorithm smaller than its original length, being no law (algorithm) to describe it.

Using the $K(.)$ function to measure the algorithmic complexity of $x$ and $x'$, we can achieve a measure of universal similarity through the conditional algorithmic complexity [Chaitin 2007]. Intuitively, two world-states can be considered similar if little effort is needed to transform $x$ into $x'$. Thus, using the same assumption that world-states can be modeled as outputs of a Turing machine, conditional algorithmic complexity measures the complexity of one binary sequence ($x$) given another binary sequence ($x'$). Thus:

$$S(x|x') := \frac{max\{K(x|x'), K(x'|x)\}}{max\{K(x), K(x')\}}$$

Given the assumption that there is no effort to make a world-state in itself, $K(x|\, x) \approx 0$, therefore, $S(x|\, x) \approx 0$. If there is no similarity between $x$ e $x'$, then $K(x|\, x') \approx K(x)$ and $(x'|\, x) \approx K(x')$, then, $S(x|x') \approx \frac{K(x)}{K(x)} = \frac{K(x')}{K(x')} \approx 1$. As the similarity between $x$ and $x'$ is closer to zero, more similar both world-states are. We can define that the output of our similarity function $S(.)$, should be expressed as a Real number between 0 and 1. Thus, let us imagine that we are cogitating three counterfactual world-states, and we want to transform the distance of these possible worlds ($x'$, $x''$, $x'''$) defined by $S(.)$, into probabilities, to use these values as a prior for Bayesian updating[2]. Let's say that the values of $x', x''$ e $x'''$ are respectively:

---

[2] One of the limitations of the probabilistic method of counterfactual analysis is that there is no clear way to define probabilities of actions that an agent did not perform, and therefore has a 0 probability. When we use Bayes' theorem and try

1) 0.81893085;
2) 0.54768653;
3) 0.14973508.

Each of the three numbers represents the distance assigned by $S(.)$ from the real world-state. Ideally, the world-state that has the highest distance should be converted to the world-state with the lowest probability (more distant = low probability = less vaguely true). To convert these numbers to probabilities, a CDF (Cumulative Distribution Function) can guarantee this result. We use an exponential function as an example below:

$$P(x) = \exp(-x)$$

This function gives us back the values $P(x') = 0.44090279382$, $P(x'') = 0.5782861116$ and $P(x''') = 0.8609360253$, if we divide each of the values by the sum of the three, we get a normalized probability distribution, $P(x') = 0.23450718$, $P(x'') = 0.30757856$ and $P(x''') = 0.45791426$, which when added together result in 1, satisfying Kolgomorov's 2nd axiom. Thus, besides serving as a metric between world-states, such a

---

to condition the probability of an event (A) to an action (B) that did not occur, and therefore has zero probability, we end up with an indefinite result: $\text{Probability of } P(A|B) = \frac{P(B|A)P(A)}{P(B)}, \text{if } P(B) = 0, \text{then } P(A|B) \text{ is undefined}$. Thus, in a way, the probabilistic analysis method, to be valid, requires the agent to be omniscient, something unrealistic. Thus, in principle, the method proposed above serves as a hyperprior for the stipulation of uncertain probabilities.

function provides conditional probabilities to the real world-state of the agent.

Now, we need to define whether $x$ and $x'$, the real and the counterfactual world-states, are produced by the same environment $w$. That is, we want to avoid cases of possible worlds that look *extremely similar to the real world, and yet, are governed by totally different physical laws.*

For example, 'what is the distance between the real world and the possible world where everything is the same, but I am a Wizard?' [3] In this way, we do not want $S(.)$ to falsely attribute a low distance to this kind of possible world. We again assume that the least complex environment, given the representation of the world-state $x$, is the one that produces $x$. This principle can be defined as Ockham's razor, or in algorithmic and mathematical terms, Solomonoff's Universal distribution. Through this tool, we can find within the probability distribution of possible computable environments, $W$, the environment e most likely to produce the world-state $x$.

The algorithmic probability distribution over possible environments is defined by $2^{-K\,(w)}$. Thus, given a world-state $x$, this algorithmic probability distribution will assign a high probability to simpler environments, which could produce $x$, because the simplicity of an algorithm is inversely proportional to the size of the program that computes it. Thus, to determine which is the simpler environment $w$ would produce $x$, we use the following function:

$$SI(x) \coloneqq \max_{w} \sum_{w \in W} 2^{-K(w)}$$

[3] A wizard here meaning that the speaker could break the laws of physics at will.

Where $SI(x)$ is the environment most likely to compute $x$, the real/counterfactual world-state proposed, and $W$ is the set of all computable environments that would produce $x$, relative to the reference Turing machine $U$. Given the probability distribution $2^{-K\,(w)}$ summed overall environments $w \epsilon W$, $\underset{w}{max}$ will result in the environment $w$ with the greatest probability, the simplest, of producing the sequence $x$. If the environment $w$ is the same that produced both $x$ and $x'$ sequences, then:

$$\Delta SI(x,x') = \frac{SI(x)}{SI(x')} \approx 0$$

Different environments where the intrinsic dynamics are different, even if they produce a counterfactual world-state $x'$ conditionally similar to the real world-state $x$, would have a higher difference in the algorithmic probability distribution $\Delta SI(.)$. Thus, we arrive at two functions to estimate the similarity between possible worlds:

- $S(x|x')$ establishes the *similarity* between possible world-states;
- $\Delta SI(x,x')$ measures the difference between the internal dynamics of the real world and the counterfactual world, prioritizing *simplicity*.

With these functions, we prioritize possible worlds that resemble more the real world, and at the same time, possible worlds that are governed by the same laws that govern our world. In case an environment cannot be compressed into a smaller algorithm than its complete description, then this particular $w$ will only generate stochastic world-states, and cannot be understood or predicted, being only *'what it is'*.
We can understand the type of analysis suggested also in a heuristic way. Agents, when engaging in imaginative reasoning or counterfactual oratory, determine which propositions are more vaguely true (more similar to the real world) using two intuitive principles. The first is similarity, how similar the counterfactual scenario is to the real world-state. The second is simplicity, how different the real world-state should be for such a counterfactual world-state to be possible. The suggested idea is that counterfactual scenarios closer to the real world-state have less conditional algorithmic complexity, while more complex scenarios are more distant. This similarity can be measured by the algorithmic complexity of the world-state in question.

## 6. A dialogue with the literature

First, the restrictions of the proposed model are:

- *Strong Centralisation:* the concept of similarity motivates the following idea. If $w$ is already a $\phi$-world, then the $\phi$-world most similar to $w$ is $w$ itself. There is no effort to make a world-state in itself, $K(x|\ x) \approx 0$;
- *Plurality:* there is not always a single $\phi$-world closer when evaluating a possible counterfactual world-state $\phi > \psi$, because when the differences between world-states are determined by random variables, no form of compression is possible, and world-states should have a uniform distance and probability distribution;

- *Limit Assumption:* as you move into $\phi$-worlds closer to the $w_0$, you reach a limit determined by random variables, and you cannot reach a $\phi$-world closer to $w_0$.

While we can say that both Stalnaker and Lewis agree on the principle of Strong Centralisation, their views differ on the Singularity and Limit assumptions. While Stalnaker [1968] endorses Singularity, Lewis [1973b] favours Plurality of possible worlds.

According to the proposed model, random sequences cannot be compressed into a smaller algorithm than their description. After all, something that cannot be expressed by any pattern is the very definition of randomness. Thus, when possible world states differ by random variables, such as the playing of a fair coin, a lottery, the decay of a radioactive atomic nucleus, or quantum fluctuations, there is no point in questioning which world is the most similar. After all, to accomplish such a task we would need to be able to compress randomness, that is, to describe it in a form smaller than its total definition, and this leads us to a contradiction.

Regarding the Limit Assumption, we can argue that a corollary result of our outcome regarding the Plurality assumption is that the limit of proximity between world-states is also limited by random variables. Thus, when the differences between a possible counterfactual world, and the real world, are restricted to only matters represented by stochastic variables, this is the limit.

Again, for there to be a continuum of possible worlds ever closer together we would need a way of compressing more and more random sequences, which leads us to the same contradiction.

For Stalnaker [1968], there should be a chain of possible worlds that leads to the closest possible world (Singularity).

However, as we argue, this chain can lead to several equally close possible worlds (Plurality).
Lewis [1973] argues that there is no limit so that possible worlds resemble the real world continuously and without limit. However, according to the argument just explained, this proposition also leads us to a contradiction. Another corollary result would be that the Law of excluded middle, as in the example of playing a fair coin used in Session 3, would not apply to counterfactuals involving random variables. In the case of a fair coin toss, both possible worlds (head and tails) are equally close to the real world.
The algorithmic complexity analysis model is best understood when dealing with situations where the distance between possible worlds depends on a stochastic source component. Thus, we'll present an example involving lotteries to better illustrate the arguments just explained:

- The Binary Lottery is a fair lottery that draws five random numbers from a sample of 256 numbers. Ana has a lottery ticket with the numbers [01000111], [00101011], [01000010], [01010111] and [01100011]. Respectively, 71, 43, 66, 87, and 99. However, the numbers drawn on lottery day were 71, 43, 66, 87, and 100. How far from the real world is the possible world where everything is the same, but the only difference is the last winning lottery number. What if instead of raffling the number 100, the number 99 had been raffled?

If the lottery is a fair one, then the result of the five numbers raffled is *algorithmically incompressible.* No algorithm can predict the next digit. Thus, in statistical terms, the raffle must have a uniform probability distribution between 0 and 256. The current world $x$, [71, 43, 66, 87, 100], is equally distant from all possible lottery results where the last number raffled is different. There are 252 possibilities, 252 possible worlds equally distant from $w_0$, and in only one of them ($x'$= [71, 43, 66, 87, 100]) Ana is the lottery winner. If we evaluate each of the numbers raffled, through their binary representation using the same function that we showed in the previous session, $P(x) = exp(-x)$.
After calculating the value of each of the numbers raffled we get the standardized results listed below:

- $P(01000111) = 0.990048734794 \approx 20\%$
- $P(00101011) = 0.998990399989 \approx 20\%$
- $P(01000010) = 0.990049734744 \approx 20\%$
- $P(01010111) = 0.989949734871 \approx 20\%$
- $P(01100011) = 0.989060169979 \approx 20\%$

The above result follows our intuitions, where a sequence of random numbers has a uniform probability distribution. If we had done the same procedure for all 252 numbers, the result would be the same for all numbers, $\frac{1}{252}$ = 0.004%. This shows how in cases where we have algorithmically incompressible information, all possible counterfactual worlds are equally close. So even if the probability that Ana won the lottery is very low, all possible worlds are equally, incompressibly, distant. Perhaps the only way to counter this example would be to demonstrate a way to predict random numbers, which amounts to predicting lottery results!
In response to the criticisms made towards the similarity analysis method, reviewed in section 4 [Morreau, 2010, Hájek, 2014], we argue that there are formal notions of similarity used in science. Even though the description made by Lewis and Stalnaker is, in fact, informal, we can formalize the similarity analysis method using algorithmic complexity. As seen above, this type of analysis brings a new light to impasses concerning the Singularity and Limit assumptions, and as a bonus, even probabilistic priors can be estimated. Now, concerning Goodman's problems, vagueness, context-dependence, and non-monotonicity, the method proposed resembles the semantics of possible worlds in trying to accommodate them. With vagueness, in the same way as similarity and probabilistic analysis, we argue that the truth condition of a counterfactual diminishes as it distances itself, in similarity and simplicity, from the real world. Thus, the propositions:

a) If that piece of butter had been heated to 150º, it would have melted.
b) If this piece of butter had been heated to 150º, it would not have melted.

Are not equally true. The simplicity function $\Delta SI\,(.)$ attributes a greater probability to possible worlds more similar to the present speaker's world, so that, by preserving the physical laws of our universe (let us suppose that the speaker lives in our universe), (a) is more vaguely true than (b). Now, concerning context-dependence, responding to the criticism of Quine [1960; 1982], who stated that it would be impossible to determine which proposals would be more true:

c) If Caesar had been in command of Korea, he would have used the atomic bomb.
d) If Caesar had been in command of Korea, he would have used catapults.

For Quine, the impossibility would arise from the fact that counterfactual world-states have no objective grounding, and are linked only to the imagination and purpose of the speaker.

We agree that counterfactuals are, in fact, creations of an agent's imaginative reasoning. However, they can be grounded in objective real-world premises, and this is the assumption of Strong Centralization. The objective grounding provides a reference to the counterfactual, the current world of the speaker ($w_0$).

In the proposed model, both the $S(.)$ function and $\Delta SI\ (.)$ formalize the idea that possible worlds more similar and simpler should be more vaguely true. What we have in the example of Quine would not be a dependence on context, but a lack of information. In other words, Quine does not give us the speaker's central point, his $w_0$.

If Caesar were in charge of Korea in 40 BC, then (d) is more vaguely true than (c) (assuming that a world where atomic weapons were invented before gunpowder would be very different from ours). While Caesar was in command of Korea between 2006 and 2009, then (c) is more vaguely true than (d) (assuming that the world most similar to ours is the one in which Caesar is a good military strategist, not a completely insane one).

The last problem concerns the non-monotonicity of counterfactual statements:

e) If I had struck this match, it would have lit.
f) If I had struck this match and done so in a room without oxygen, it would have lit.

For Goodman, the main problem of non-monotonicity is that it is quite difficult to specify all the details and background factors, together with the physical laws in force, in non-counterfactual semantic terms. We agree with Goodman that counterfactuals should be by definition non-monotonic and that a full semantic definition of the real world would be intractable.

However, the proposed method does not use semantics to assess similarity and conditional probability between possible worlds but rather the information contained in the real and counterfactual world-states. The conversion of a real/possible world to a digital world-state, where we represent information by bits, allows a great simplification of the specification problem.
For example, a complete definition of the laws of physics in semantic terms is much more complex than its algorithmic specification. We can specify such laws succinctly in algorithms, and that is what allows contemporary physics to work with computational simulations to explore phenomena that would be, on the contrary, unobservable.
In more intuitive terms, an agent embedded in the environment, who knows the internal dynamics of the environment and is capable of counterfactual reasoning, can counterfactually perceive violations of these principles. What Solomonoff's universal distribution gives us is a way to assign a greater probability to possible worlds which are more similar to the real world. Worlds with the fewest possible violations. Thus, defining without the use of semantics, the intrinsic dynamic of the environment.

## 7. Limitations

What are the limitations of the proposed method?
First, our model allows for a notion of universal similarity between all possible computable worlds, and the first criticism that we'll point is that this is too general.
In other words, we make the space of possibilities so great, so vast, that any kind of investigation (through all possible computable worlds) would be intractable. The only physical and metaphysical restriction made is that the laws of our universe must be preserved, in their simplest form, and that we delimit the metaphysical space to only computable worlds. However, is the reductionism of digital physicalism acceptable? Would physicality be anything other than computability? Whether the universe or the multiverse can be adequately modeled as a Turing machine remains an open question.

Perhaps the greatest obstacle (in terms of practical implementation) is the fact that algorithmic complexity is in itself an incomputable function, meaning that there is no general algorithm that can attest to the algorithmic complexity for any finite sequence $x$.

This may seem extremely intuitive. Otherwise, all the infinite (finite in length) sequences of possible bits, even the random ones, could be generated by a single finite program, $K(.)$, with a complexity smaller than $x$ bits. How could a program of length $n$ generate random sequences of length $n + 1$? Another contradiction.

Even so, the practical impossibility does not invalidate the theoretical insights acquired, especially the ones concerning how world-states differentiated by random variables should behave respecting principles like simplicity, complexity, and plurality.

## 8. Conclusion

In this study, we would like to make clear two important points regarding two different methods of counterfactual analysis. While Lewis-Stalnaker's Possible worlds semantics fails to provide a more formal and rigorous similarity function for estimating the distance between possible worlds, Adams' probabilistic analysis model fails to deal with situations where the antecedent probability distribution is unknown. A well-known Achilles heel of the Bayesian paradigm.

In this proposal, we sought to develop a similarity function for counterfactual analysis, inspired by the ideas of Lewis and Stalnaker and made possible by tools of Algorithmic Information Theory. At the same time, the model proposed provides an objective basis for the estimation of conditional probabilities, and it more rigorously formalizes the concept of similarity between possible worlds. Furthermore, using Solomonoff's universal distribution, we extend the concept of similarity to that of simplicity. We believe that these guiding principles are both intuitive and philosophically justifiable.

Even provided the incomputability of algorithmic complexity, the proposed methodology was able to dialogue with the existing literature, shedding new light on the debate between Lewis and Stalnaker regarding the Singularity and Limit assumptions. Together, we sought to answer criticisms and questions raised by the literature against Possible world semantics and similarity analysis methods in general. As a final message, this study had as one of its main motivations to show how tools from other areas can come to assist in philosophical investigations. When an object of study becomes of interest to a large number of different fields of knowledge, interdisciplinary research must be sought. Who knows what we may discover when we import tools from another box.